# Panic-induced symmetry breaking in escaping ants

E. Altshuler, O. Ramos, Y. Nuñez & J. Fernández

Henri Poincaré Group of Complex Systems & Superconductivity Laboratory,

Physics Faculty-IMRE, University of Havana, 10400 Havana, Cuba

Helbing *et al*. recently made an important step forward in a theoretical paper where a detailed simulation of individuals escaping from a closed room was presented[1]. This work has been rapidly followed by other theoretical and experimental research[2-7]. One of the most unexpected phenomena predicted by Helbing *et al.* is the symmetry breaking in the escape from a room with two identical exits. Here we report experiments on ants that corroborate this behavior and show that the escape dynamics of ants under controlled panic can serve as a model to study humans in analogous situations.

In a first type of experiments (Experiment I), we introduced a group of approximately 80 individuals of the ant species *Atta insularis*[8,9] collected from natural nests, into a circular cell covered by a glass plate with two exits symmetrically situated at left and right (Fig. 1a), which were initially blocked. Then, we opened the exits synchronously, and counted the number of ants abandoning the cell through each exit until it was empty. Fig. 1b shows a graph describing quantitatively the result of a sample run of Experiment I. Although the difference in the use of the two doors eventually reaches a value of 15 ants at a certain stage of the run (roughly corresponding to 20% of the total of escaping ants), it becomes clear that both doors have been used almost symmetrically at the end.

A very different output, however, results from a second kind of experiments (Experiment II). In this case everything takes place as in Experiment I, with the important difference that, a few seconds before opening the doors, a dose of 25 or 50 µl of an insect repelling liquid is rapidly injected in the cell through a hole in the covering glass, producing a disk-shaped spot of the substance at the center of the filtering paper on which the whole setup rests (Fig 1c). A sample output of Experiment II is shown in Fig. 1d. Differently from Experiment I, one of the doors is always much more used to escape than the other one, which can be described as a symmetry breaking induced by

"panic" associated to the repelling agent. In the specific run depicted in Fig 1c,d the difference in the use of the two doors reached nearly 50 ants at the end of the experiment (around 60% of the total number of escaping ants). Besides symmetry breaking, a second difference between Experiment I and II clearly seen in Fig b, d is that, in the latter, the total time of escape is much smaller, probably because the "desired velocity"[1] of the ants increases due to the effect of the repelling agent.

Similar results were observed in several runs of Experiment I and II: statistics showed that the difference in use between the two exits at the end of the experiment were, in average, 12% ± 3% for Experiment I and 51% ± 7% for Experiment II. The "preferred exit" in Experiment II was either the left or the right one, with no connection to any source of assymetry in the experimental setup or to the spatial distribution of ants inside the cell before opening the doors. The experimental results were also quite similar when using ants collected from the same nest, or from different nests no more than 20 meters apart from each other. They were also similar when repeated on the same group of ants.

These results are coherent with the theoretical predictions reported by Helbing *et al.*[1] They defined a "panic parameter" which induces individualistic behaviour (each pedestrian tends to find an exit by him/herself) when low, and herding behaviour (pedestrians tend to follow the crowd) when high. In their two-exit room simulations, the authors find that a high value of the panic parameter produces jamming at one of the doors, thus provoking inefficient escape. This tendency, illustrated in Fig 3a of reference 1 is analogous to our Fig 1c, which shows that their panic parameter is related to the effect of the repelling substance used in our experiments.

In spite of the huge behavioural differences between humans and ants in normal conditions[10], our experiments suggest that some features of the collective behaviour of both species can be strikingly similar when escaping under panic.

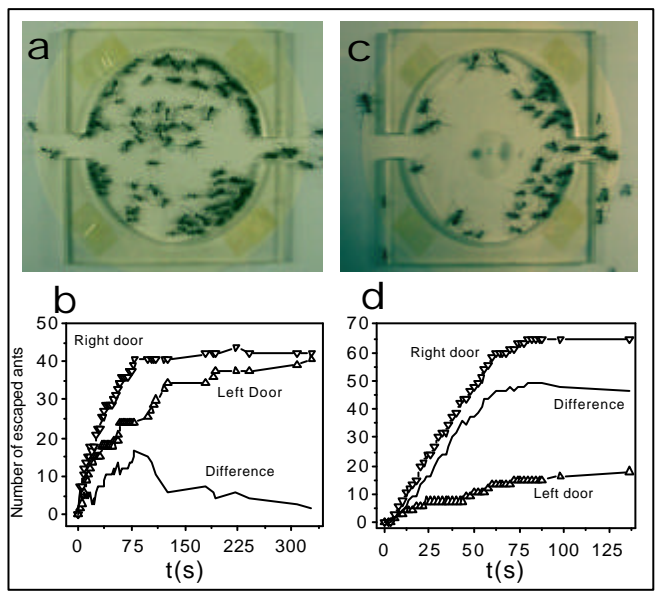

**Figure 1** Escape of ants from a cell with two symmetrically located exits. The cell consists in an acrylic drum of 80 mm diameter and 5 mm height, with two exits 10 mm wide symmetrically situated at left and right positions. The drum rests on an circular piece of filtering paper laying on an horizontal surface, and is covered by a flat glass plate of 3 mm thickness with a hole of 2 mm diameter situated at the center of the drum. Approximately 80 ants from the species *Atta insularis* are added with both exits blocked by acrylic bars. **a,** ant distribution several seconds after opening the exits at $t = 0$. **b,** number of ants abandoning the cell as time goes by. **c,** ant distribution several seconds after adding 50 µl of an insect repellent fluid (*Citronela*, Labiofam, Cuba) though the central hole, and then opening the exits at $t = 0$ (note a circular spot of repellent fluid of approximately 20 mm diameter at the centre of the cell). Clogging of ants at the right exit is clearly visible. **d,** number of ants abandoning the cell as time goes by, quantitatively demonstrating the symmetry breaking when repellent is added. Small deviations from the ideal circular shape of the repellent spot or from its location at the center of the setup did not produce important effects in the experimental output.